\magnification=\magstephalf
\input amstex
\loadbold
\documentstyle{amsppt}
\refstyle{A}
\NoBlackBoxes

\vsize=7.5in

\def\pf{\hfill $\square$}
\def\c{\cite}
\def\a{\alpha}
\def\b{\beta}

\def\fg{\frak{g}}
\def\fh{\frak{h}}

\def\end{\text{End}}

\def\p_x{\partial_{x}}

\topmatter
\title  Liouville integrability of a class of integrable spin Calogero-Moser 
systems and exponents of simple Lie algebras
  \endtitle
\leftheadtext{L.-C. Li, Z. Nie}
\rightheadtext{Liouville integrability of integrable spin CM systems}
\author Luen-Chau Li and Zhaohu Nie\endauthor
\address{L.-C. Li, Department of Mathematics, Pennsylvania State University,
University Park, PA  16802, USA}\endaddress
\email luenli\@math.psu.edu\endemail
\address{Z. Nie, Department of Mathematics, Pennsylvania State University,
Altoona Campus, 3000 Ivyside Park, Altoona, PA 16601, USA}\endaddress
\email zxn2\@psu.edu\endemail
\abstract  In previous work, we introduced a class of integrable spin 
Calogero-Moser systems associated with the classical dynamical r-matrices 
with spectral parameter, as classified by Etingof and Varchenko for
simple Lie algebras.  Here the main purpose is to establish the
Liouville integrability of these systems by a uniform method.

\endabstract
\endtopmatter

\document
\subhead
1. \ Introduction
\endsubhead

\baselineskip 15pt
\bigskip

Systems of spin Calogero-Moser (CM) type are Hamiltonian systems with very rich
structures.   After the initial example of Gibbons and Hermsen \c{GH},
a variety of such systems have appeared in the literature over the years. (See, for example, 
\c{BAB1, BAB2, FP, HH, L1, L3, LX1, LX2, MP, Pech, P, Wo, Y}    and the references therein.) 
This is a testimonial to the relevance of such systems in various areas of mathematics
and physics.  In \c{LX1, LX2}, as a by-product of an effort to understand conceptually the
calculations in \c{BAB1, BAB2}, we introduced a class of spin CM systems associated with the classical dynamical r-matrices with spectral parameter, as defined and classified
by Etingof and Varchencko for complex simple Lie algebras
 \c{EV}.
The classical dynamical r-matrices with spectral parameter in \c{EV} are solutions of 
the classical dynamical Yang-Baxter equation (CDYBE) with spectral parameter, which was
first introduced and studied by Felder \c{F}.   While Felder studied CDYBE in the context of 
conformal field theory, we showed how to make use of the solutions of this equation to construct
and to study our spin systems.   Indeed, in \c{L2}, we showed how to obtain
the explicit solutions of the integrable spin CM systems in \c{LX2}  by using the factorization
method developed in \c{L1}.   That this is possible is due to some
remarkable geometric structures underlying the so-called 
modified dynamical Yang-Baxter equation (mDYBE) \c{L1}.   This work is a sequel to
\c{LX2} and \c{L2}.  Our main purpose here is to establish the Liouville integrability of the
integrable spin CM systems in \c{LX2} on generic symplectic leaves.

The spin CM systems constructed in \c{LX1, LX2}  are of 
three types: rational, trigonometric, and elliptic, as in the
case of their spinless counterparts in \c{OP}.  In the rational case, recall
that we have a family of rational spin CM systems parametrized by subsets
$\Delta^{\prime}\subset \Delta$ which are closed with respect to addition
and multiplication by $-1.$  Here $\Delta$ is the root system
associated with a complex simple Lie algebra $\fg$ and a fixed Cartan
subalgebra $\fh$ of $\fg.$   In the trigonometric case, there
is also a family but now the systems are parametrized by
subsets $\pi^{\prime}$ of a fixed simple system $\pi\subset \Delta.$
Finally we have an elliptic spin CM system for each complex simple Lie
algebra.  In \c{LX1,LX2}, generalized Lax operators taking values in the dual
bundles $A\Gamma$ of the corresponding coboundary dynamical Lie
algebroids $A^*\Gamma$ were constructed for these systems. 
If we let $H$ denote a connected Lie group corresponding to $\fh,$ then 
recall that in each case, the phase space $P$ of the spin CM system
is a Hamiltonian $H$-space (with equivariant momentum map
$J$) which admits an $H$-equivariant realization in the
corresponding $A\Gamma$ and the Hamiltonian is the pullback
of a natural invariant function on $A\Gamma$ under the realization
map.   It is a characteristic of these
systems that the pullback of natural invariant functions
on $A\Gamma$ to $P$ do not Poisson commute everywhere,
but they do so on $J^{-1}(0)$ in all cases.  Hence we can obtain
the integrable spin systems on $J^{-1}(0)/H$ by Poisson
reduction.  It should be pointed out in our setup, the Lax operator 
$L$ is only part of the generalized Lax operator. Indeed, for the rational 
(resp.~ trigonometric) case, if $\Delta^{\prime} \neq\Delta$ 
(resp.~ $<\pi^{\prime}>\neq\Delta$),
 the equation of motion for $L$ only carries partial
information about the dynamics and it is necessary to obtain
the missing piece from the other parts of the 
generalized Lax operator \c{L2}.  Nevertheless, as the reader will
see, the Lax operator suffices when we consider Liouville
integrability. 

The method we use to establish the Liouville integrability of 
the integrable spin CM systems can be explained as follows.
Let us begin with a familiar situation.
In the usual classical $R$-matrix theory, it is well-known that if 
the Lax operator $L$ of a finte-dimensional system
takes values in a loop algebra $L\fg,$ (i.e., $L$ depends on a 
spectral parameter), then one can
obtain integrals in involution by pulling back the ad-invariant functions
on $L\fg$ using $L.$  However, due to the finite-dimensionality
of the system,  we cannot
expect the integrals obtained in this way to be functionally
independent or nonzero even though the ring of ad-invariant functions
of $L\fg$ has an infinite number of generators.
In the case when $\fg \subset gl(n,\Bbb C)$ for some $n,$ say, then of
course there is a 
standard way to construct a finite collection of Poisson commuting integrals.
Namely, one simply writes down the characteristic polynomial of $L(z)$ and in 
this case, the completeness of the integrals can be addressed by
algebro-geometric means provided certain additional conditions are
satisfied \c{RSTS}.
(For an example where there exists a family of additional integrals besides
the ones given by the characteristic polynomial of $L(z)$, see \c{DLT}.)
In our case, we can of course take concrete representations of the
simple Lie algebras, however, an intrinsic way to construct
the necessary integrals which works for all simple Lie algebras
is clearly preferred.  What we essentially do in this work is to
substitute the elementary symmetric functions in the matrix case by the
primitive invariant polynomials of Chevalley, which are homogeneous
with degrees related to the exponents of the simple Lie algebras.
(See, for example, \c{C, K2, V}.)  To obtain the quantities
of interest, we simply evaluate the primitive invariants on 
the Lax operator and these give Poisson commuting integrals
on $J^{-1}(0)$ by the general theory in \c{LX2, L1}.
To count the number of nontrivial integrals obtained
in this manner, our basic realization is that we can appeal to
a theorem of Shephard and Todd \c{ST}
relating the sum of exponents of a complex simple Lie algebra to the number
of roots of $(\fg,\fh).$   Of course, there remains the task
of showing that the nontrivial integrals are functionally
independent on an open, dense set of the phase space.
As the reader will see, we can also accomplish this in a
uniform way due to some common structure which exists
among the three types of spin CM systems.  To conclude, we
remark that the method which we develop here to
construct and count the number of integrals is a 
general method.  In principle, it should work for other systems associated
with simple Lie algebras and with spectral parameter dependent
Lax operators.  Thus what we show in this work is just an
illustration of this general method. Furthermore, some
of our analysis involving invariant polynomials
(see Lemma 3.1 and Theorem 6.4) may also be of independent 
interest in Lie theory.

The paper is organized as follows.  In Section 2, we present
for the most part some background material for the reader's
convenience, we also take the opportunity to set up the notations.
In the first subsection, we begin by summarizing some basic
facts about the invariant polynomials of Chevalley and the
exponents of simple Lie algebras which are of relevance here. 
We also recall some of the tools which we find useful
in dealing with these invariant polynomials. (Further tools
will be developed in subsequent sections.)  In the
second subsection, we recall the construction of the 
class of spin Calogero-Moser systems associated with the
classical dynamical r-matrices with spectral parameter.
We then explain how Poisson reduction gives rise to the
associated integrable models.  At the end of the subsection,
we conclude with our first result, namely, the connection
between the dimension of the maximal dimensional phase
spaces of our systems and the exponents of the complex simple Lie algebras.
In Section 3, we construct the integrals for the rational
case by evaluating the primitive invariants on the Lax
operators and we count the number of nontrivial integrals.
As it turns out, for each primitive invariant $I_k$, exactly one
of the quantities which arise in the expansion of $I_k(L(z))$
in $z$ is identically zero in this case while another one is a 
Casimir function. In Section 4 and 5, we do the same for the 
trigonometric case and the elliptic case.  Finally, in Section 6, we show 
that the integrals constructed in Sections 3-5 are functionally
independent, thus proving the Liouville integrability of the
systems on generic symplectic leaves.

\bigskip
\bigskip

\subhead
2. \ Preliminaries
\endsubhead
\bigskip

In \c{LX2}, we introduced a class of integrable spin Calogero-Moser
systems associated with the classical dynamical r-matrices with
spectral parameter, as classified by Etingof and Varchenko \c{EV}
for complex simple Lie algebras.  Our goal in this section is to establish 
Proposition 2.2.6 which gives the dimension of the maximal
dimensional phase spaces of such systems in terms of the exponents
of the complex simple Lie algebras.  For the reader's convenience, we will 
provide some background material, we will also take the opportunity to
set up the notations.   In the first subsection, we will begin by summarizing a 
number of basic facts about the  invariant polynomials of Chevalley and the
exponents of simple Lie algebras. We will also collect here some of the tools 
which we find useful in dealing with these polynomials.  In the second 
subsection,
we will recall the class of integrable spin Calogero-Moser systems in \c{LX2}.
Then we will present our first result which we alluded to above, thus tying
together the two subsections.

\subhead
2.1 \ The invariant polynomials of Chevalley and the exponents 
\endsubhead
\medskip

Let $\fg$ be a complex semisimple Lie algebra of rank $N,$
and let $G$ be a connected Lie group with $Lie(G) = \fg.$
We recall that the group $G$ acts on the algebra ${\Cal P}(\fg)$ 
of polynomial functions on $\fg$ by
$g\cdot P = P^g$,  $g\in G$, $P\in {\Cal P}(\fg)$, where
$$P^g(x) = P(Ad_{g^{-1}} x), \,\, x\in \fg.\eqno(2.1.1)$$
Let $I(\fg)$ denote the ring of polynomial functions on
$\fg$ invariant under the above action of $G$.  Then the
well-known theorem of Chevalley \c{C} asserts that $I(\fg)$ is generated by
$N$ algebraically independent homogeneous polynomials 
$I_1 ,\ldots,I_N$.  In other words, if ${\Bbb C}[Y_1,\ldots,Y_N]$
denotes the polynomial ring in the $N$ variables $Y_1,\ldots, Y_N$,
then 
$$I(\fg) = {\Bbb C}[I_1,\ldots,I_N].\eqno(2.1.2)$$
Let us denote by $\fh$ a fixed Cartan subalgebra of $\fg,$
and let $W$ be the Weyl group of the pair $(\fg,\fh)$
generated by reflections in the hyperplanes in $\fh.$
Then indeed the restriction of $I(\fg)$ to $\fh$ is an algebra
isomorphism of $I(\fg)$ onto the algebra ${\Cal P}(\fh)^{W}$
of polynomials on $\fh$ which are invariant under $W.$
Write
$$\deg I_{k} = d_{k},\quad k=1,\ldots, N. \eqno(2.1.3)$$
We will assume that the $I_{k}$'s are ordered in the sense
that
$$d_1\leq d_2\leq\ldots\leq d_N.\eqno(2.1.4)$$
Following Kostant \c{K2}, we will refer to the $I_k$'s as the
primitive invariants.  The numbers $m_k= d_k -1$, $k=1,\ldots, N$,
are called the exponents of $\fg$ and are the basic invariants
of $\fg$ \c{B}. (See also \c{CM} and the references therein.)
For the purpose in this work, we will need the following 
results due to Shephard and Todd \c{ST}.

\proclaim
{Theorem 2.1.1 \c{ST}}  Let $m_k$ be the exponents of $\fg,$ $k=1,\cdots, N.$
Then 
$$\eqalign{\sum_{k=1}^{N} m_{k} & =  \# \,\,\hbox{of reflections in}\,\, W\cr
                             & = {1\over 2} (\#\,\, \hbox{of roots of}\,\,
                                  (\fg, \fh))\cr
                             &  = {1\over 2}(\dim \fg - N).\cr}\eqno(2.1.5)$$
\endproclaim

As the reader will see, (2.1.5) is crucial
in counting the total number of integrals which
we construct by evaluating the primitive invariants on
the Lax operators of the integrable spin Calogero-Moser systems.  
We will explain this in the next subsection below.

While the problem of computing the exponents was originally motivated
by the problem of computing the Betti numbers of complex simple
Lie groups, it turns out that there is a different way to describe
these numbers which is relevant for us.  For this purpose, let
us introduce some notation.  First of all, we will assume from
now on that $\fg$ is simple with Cartan sublagebra $\fh$ and
Killing form $(\cdot,\cdot),$ and
let $\fg = \fh \oplus \sum_{\alpha\in \Delta} \fg_{\alpha}$ be
the root space decomposition of $\fg$ with respect to $\fh.$
We fix a simple system of roots
$\pi =\{\alpha_1,\ldots,\alpha_N\}$, and denote by $\Delta^{\pm}$
the corresponding positive/negative system.   If $\alpha\in \Delta^{+},$
recall that we can express $\alpha$ uniquely as a sum of simple root 
$\sum_{i=1}^{N} n_{i} \alpha_{i},$  where $n_i$ are non-negative
integers.  The height of $\alpha$ is defined to be
the number
$$\hbox{ht}\,(\alpha) = \sum_{i=1}^{N} n_i.\eqno(2.1.6)$$

\proclaim
{Theorem 2.1.2 \c{K1}} If $b_j$ is the number of $\alpha\in \Delta^{+}$ 
such that $\hbox{ht}\,(\alpha) =j,$ then
\newline
(a) $b_j -b_{j+1}$ is the number of times $j$ appears as an
exponent of $\fg,$
\newline
(b) $N = b_1\geq b_2\geq \cdots \geq b_{h-1} =1,$
where $h$ is the Coxeter number.  Moreover, the partition of
$r = |\Delta^{+}|$ as defined by the above sequence of numbers
is conjugate to the partition 
$h-1=m_{N}\geq m_{N-1}\geq \cdots \geq m_{1} =1.$
\endproclaim 

In the rest of the subsection, we will summarize a number of
basic facts from \c{K2} that we will use in Section 3 and
more significantly in Section 6 below.
To begin with, we recall that the symmetric algebra
$S = S(\fg^*)$ can be identified with ${\Cal P}(\fg).$
On the other hand, we can associate to each $x\in \fg$ a differential
operator $\partial_{x}$ on $\fg,$ defined by 
$$\partial_{x}f(y) = {d\over dt}{\Big|_{t=0}} f(y + tx), \quad f\in
C^{\infty}(\fg).\eqno(2.1.7)$$
In this way we have a linear map $x\mapsto \partial_{x}$ which can be
extended to an isomorphism from the symmetric algebra
$S_{*} = S(\fg)$ to the algebra of differential operators
$\partial$ with constant coefficients on $\fg.$  From now onwards we
will identify the two spaces and with this identification,
we have a nondegenerate pairing between $S_{*}$ and $S$
given by
$$
\langle \partial,f\rangle=\partial f(0),\eqno(2.1.8).
$$
where $\partial\in S_{*}$, $f\in S,$ and $\partial f (0)$ denotes
the value of the function $\partial f$ at $0\in \fg.$
It is obvious that both $S_*$ and $S$ are graded:
$S_{*} = \oplus_{j\geq 0} S_j$, $S =\oplus_{j\geq 0} S^{j}.$
If $f\in S^m$ and $x\in \fg,$ it follows from the Taylor expansion that
$$\Big\langle \frac{(\p_x)^m}{m!},f\Big\rangle=f(x).\eqno(2.1.9)$$
Now $S$ is a $G$-module by (2.1.1).   On the other hand, it is
clear that the adjoint
action of $G$ on $\fg$ can  be naturally extended to an action
of $G$ on $S_{*}.$   Therefore in view of (2.1.1), we have 
$$\langle g\cdot\partial, g\cdot f\rangle = \langle \partial, f\rangle,
\eqno(2.1.10)$$
for all $g\in G,$ $\partial \in S_{*}$ and $f\in S.$
By differentiation, $S$ and $S^*$ become $\fg$-modules and
the actions of $\fg$ on both spaces are by derivations.
Therefore we have the ``product rule" and the ``power rule":
$$
x\cdot (\partial\delta)=(x\cdot \partial)\delta+\partial(x\cdot \delta)
,\eqno(2.1.11)
$$
$$
x\cdot\partial^n=n \partial^{n-1}(x\cdot \partial), 
\eqno(2.1.12)
$$
for all $x\in \fg,$ $\partial,\delta\in S_*,$ and $n\in \Bbb N.$
For $y\in \fg,$ we also have
$$ x\cdot \partial_{y} = \partial_{[x,y]}.\eqno(2.1.13)$$
Since the pairing between $S_{*}$ and $S$ obeys (2.1.10),
it follows that
$$\langle x\cdot \partial,f\rangle+\langle \partial,x\cdot f\rangle=0
\eqno(2.1.14)$$
for all $x\in \fg,$ $f\in S.$  In particular, this implies that
$$\langle x\cdot \partial,f\rangle=0, \,\,\,\, \hbox{for all}\,\,f\in I(\fg)
\eqno(2.1.15)$$
since $x\cdot f =0$ for $f\in I(\fg).$

Now let $x_0$ be the unique element in $\fh$ such that
$
\a_i(x_0)=1,\ i=1,\cdots,N.
$
Then $\a(x_0)=\hbox{ht}(\a)$, and 
$
[x_0,e_{\a}]=\hbox{ht}(\a) e_{\a}
$
for all $\a\in \Delta.$  For each $j\in \Bbb Z,$ let
$$
S_*^{(j)}=\{\partial\in S_*|x_0\cdot \partial=j\partial\}.\eqno(2.1.16)
$$
Then $S_*=\sum_{j\in \Bbb Z} S_*^{(j)},$ and if $\partial\in S_{*}^{(j)},$
we will say $\partial$ has weight $j.$  
Cleary, we have
$$
\partial_{e_{\a}}\in S_*^{\left(ht\,(\a)\right)},\ \partial_p\in S_*^{(0)}
\eqno(2.1.17)
$$
for $\alpha\in \Delta,$ $p\in \fh.$   Also, 
$$S_*^{(i)}S_*^{(j)}\subseteq S_*^{(i+j)}.\eqno(2.1.18)
$$

The following consequence of (2.1.15) is very important to us in
Section 6 below. If 
$\partial\in S_*^{(j)}$ for $j\neq 0$, then 
$\partial=\frac{1}{j} x_0\cdot \partial$ and hence we have
$$\langle \partial,f\rangle =0\,\,\,\,\hbox{for all}\,\,f\in I(\fg).
\eqno(2.1.19)$$
Analogously, we let $\fg^{(j)}$ be the eigenspace of $\hbox{ad}\,x_0$ for
the eigenvalue $j.$  Then  $\fg = \oplus_{j\in \Bbb Z} \fg^{(j)}$
and 
$$[\fg^{(i)}, \fg^{(j)}]\subset \fg^{(i+j)}.\eqno(2.1.20)$$

\medskip

\subhead
2.2 \ A class of integrable spin Calogero-Moser systems and their phase 
\phantom{fake}   spaces
\endsubhead
\medskip

We recall that $\fg$ is a complex simple Lie algebra with
Cartan subalgebra $fh,$ and $\Delta^{\pm}$ are
the positive/negative system relative to a fixed
simple system $\pi$ of roots.
For each positive root $\alpha \in \Delta^+$, let 
$e_{\alpha} \in \fg_{\alpha}$ and
$e_{-\alpha} \in \fg_{-\alpha}$  be root vectors which are dual with respect to
$(\cdot, \cdot )$ so that $[e_{\alpha} , e_{-\alpha}]=H_{\alpha}$, where
the latter is the unique element in $\fh$ which corresponds to
$\alpha$ under the isomorphism induced by the Killing form 
$(\cdot,\cdot)$. We
also fix an orthonormal basis $(x_i)_{1\le i\le N}$ of $\fh$, and
write $p=\sum_{i} p_{i} x_{i}$, \, \,$\xi=\sum_{i} \xi_{i} x_{i} +
\sum_{\alpha \in \Delta} \xi_{\alpha} e_{\alpha}$ for
$p \in \fh$ and $\xi \in \fg.$  Lastly, we let $H$ be a connected Lie 
subgroup of $G$ with $Lie(H)=\fh.$

Let $r$ be a classical dynamical r-matrix with spectral parameter
in the sense of \c{EV}, with coupling constant equal to $1.$
We will fix a simply connected set $U\subset \fh$ on which $r(\cdot,z)$
is holomorphic.  By Proposition 4.5  of \c{LX2}, we can
construct the associated $H$-equivariant classical dynamical r-matrix 
$R:U\longrightarrow L(L\fg,L\fg)$, where $L\fg$ is the loop algebra
of $\fg$ and $L(L\fg,L\fg)$ is the space of linear maps on 
$L\fg.$  Indeed, it was established
in \c{LX2} that $R$ is a solution of the modified dynamical
Yang-Baxter equation (mDYBE).  Hence  we can equip 
$A^{*}\Gamma = T^{*} U \times L\fg^* \simeq TU \times L\fg$
($L\fg^*$ is the restricted dual of $L\fg$) with a Lie algebroid
structure, the so-called coboundary dynamical Lie algebroid
associated to $R.$   Our construction of the class of spin
Calogero-Moser system and its realization is based on the following
result.

\proclaim
{Theorem 2.2.1 \c{LX2}}  The map 
$\rho = (m,\tau, L): A^{*}\Omega\simeq TU \times \fg \longrightarrow 
TU\times L\fg \simeq A\Gamma$ 
given by
$$\rho (q,p,\xi) = (q, -\Pi_{\fh}\xi, p + r^{\#}_{-}(q)\xi)\eqno(2.2.1)$$
is an H-equivariant Poisson map, when the domain is equipped
with the Lie-Poisson structure corresponding to the trivial Lie
algebroid $A\Omega\simeq TU\times \fg$, and the target is equipped with
the Lie-Poisson structure corresponding to $A^*\Gamma$.  Here,
$H$ acts on $TU\times \fg$ and $TU\times L\fg$ by acting on
the second factors with the adjoint action and 
the map $r^{\#}_{-} (q) : \fg \longrightarrow L\fg$ is defined by
$$((r^{\#}_{-} (q)\xi)(z), \eta) = (r(q,z), \eta \otimes \xi) \eqno(2.2.2)$$
for $\xi$, $\eta \in \fg.$
\endproclaim
\smallskip

\remark
{Remark 2.2.2}  (a) The Lie-Poisson structure on the dual of the trivial Lie
algebroid $A\Omega\simeq TU\times \fg$ is given by 
$\{\phi, \psi\}_{A^{*}\Omega}(q,p,\xi) = (\delta_2\phi, \delta_1 \psi)
-(\delta_1\phi, \delta_2 \psi) + (\xi, [\delta\phi, \delta\psi])$ \c{L1}.
Thus the Poisson structure is a product structure, where
$T^*U\simeq TU$ is equipped with the canonical structure, and
$\fg^*\simeq\fg$ is equipped with the Lie-Poisson structure.
Moreover, the $H$-action on $TU\times \fg$ above is a canonical
action with equivariant momentum map $J:TU\times \fg\longrightarrow \fh,
(q,p,\xi)\mapsto -\Pi_{\fh} \xi,$ where $\Pi_{\fh}$ is the projection map
to $\fh$ relative to the splitting $\fg = \fh\oplus \fh^{\perp}.$
\newline
(b) As a special case of Proposition 3.1 in \c{L1}, the dual bundle $A\Gamma$ equipped
with the Lie-Poisson structure and $H$-action as defined in the above theorem is also
a Hamiltonian $H$-space.  The corresponding equivariant momentum map 
$\gamma: A\Gamma\longrightarrow \fh$ is given by the simple
formula $\gamma(q,p,X) =p.$
\newline
(c)  The map $\rho = (m,\tau, L)$ in the above theorem is to be regarded as the
generalized Lax operator of the corresponding spin CM system whose
construction we recall below.   The map $L$, on the other hand, is the 
Lax operator.
\endremark

\medskip
Let $Q$ be the quadratic function
$$Q(\xi)= {1\over 2}\oint_c (\xi(z), \xi (z))\frac{dz}{2\pi iz}\eqno(2.2.3)$$
where $c$ is a small circle around the origin.  Clearly, 
$Q$ is an ad-invariant function on $L\fg$.

\definition
{Definition 2.2.3 \c{LX2}}  Let $r$ be a classical dynamical r-matrix with
spectral parameter with coupling constant equal to 1.  Then the
Hamiltonian system on $A^{*}\Omega\simeq TU\times \fg$ 
(equipped with the Lie-Poisson
structure as in Theorem 2.2.1) generated by the $H$-invariant Hamiltonian
$${\Cal H}(q,p,\xi)= (L^{*}Q)(q,p,\xi)= 
{1\over 2}\oint_c (L(q, p, \xi)(z), L(q, p, \xi)(z))\frac{dz}{2\pi iz}
\eqno(2.2.4)$$
is called the spin Calogero-Moser system associated to $r$.
\enddefinition
\smallskip

Note that the pullback of ad-invariant functions on $L\fg$
by the Lax operator $L$ do not Poisson commute 
everywhere.   In order to construct the integrable spin CM systems, 
we have to invoke Poisson reduction \c{MR, OR}.  For this purpose,
let $Pr_i$ be the projection map onto the $i$-th factor of
$U\times \fh\times L\fg\simeq A\Gamma, i=1,2,3,$ and let
$\pi_{0}:J^{-1}(0)\longrightarrow J^{-1}(0)/H,$
$\pi_{H}: \gamma^{-1}(0)\longrightarrow \gamma^{-1}(0)/H$
 be  the canonical projections.  If $f$ is an ad-invariant function on 
 $L\fg,$ the unique function on $\gamma^{-1}(0)/H$ determined by
 $Pr_{3}^{*}f|\gamma^{-1}(0)$ will be denoted by $\bar f,$  while the
 unique function on $J^{-1}(0)/H$ determined by $L^{*}f|J^{-1}(0)$
 will be denoted by ${\Cal F}_{0}.$   Because the map $\rho$ is an 
 $H$-equivariant Poisson map, it induces a unique Poisson map
 $\widehat{\rho}: J^{-1}(0)/H\longrightarrow \gamma^{-1}(0)/H$ 
 characterized by $\pi_{H}\circ \rho|J^{-1}(0) =\widehat{\rho}\circ \pi_{0}.$
 From the various definitions, we have
 $${\Cal F}_{0} = \widehat{\rho}^{\,*} \bar f,\quad {\Cal F}_{0}\circ 
\pi_{0}= L^{*}f|J^{-1}(0).
 \eqno(2.2.5)$$
In particular, the Hamiltonian ${\Cal H}$ of the spin CM system
in (2.2.4) drops down to ${\Cal H}_0 = \widehat{\rho}^{\,*}\bar Q$
on $J^{-1}(0)/H.$
 
 \proclaim
{Theorem 2.2.4  \c{LX2, L1}}  (a) The pullback of ad-invariant functions 
on $L\fg$ by $L$ Poisson commute on $J^{-1}(0).$
\newline
(b)  Functions ${\Cal F}_{0}= \widehat{\rho}^{\,*}\bar f$ corresponding to 
ad-invariant functions $f$ on $L\fg$ Poisson commute on the reduced space 
$J^{-1}(0)/H.$
 \endproclaim
 
\remark 
{Remark 2.2.5}  (a)  The reduced spaces $J^{-1}(0)/H$ and $\gamma^{-1}(0)/H$ are
Poisson varieties in the sense of \c{OR}.
\newline
(b) The existence of the Poisson map $\widehat{\rho}$ and the formulation of 
 Theorem 2.2.4 (b) follow from general result in \c{L1}.  In \c{LX2}, we made an
 additional assumption, we also did not have $\gamma^{-1}(0)/H$ available at the
 time.  Thus the reduction picture obtained there was not an intrinsic one.
\newline
(c)  In \c{L1}, we obtain an intrinsic expression for the Lie-Poisson
structure on the dual bundle of a coboundary dynamical Lie algebroid,
from which it is clear that functions which are obtained as pullback
of ad-invariant functions under the map $Pr_3$ Poisson commute on
$\gamma^{-1}(0).$   Thus in hindsight, the result in Theorem 2.2.4 (a) is 
just a consequence of this fact, Theorem 2.2.1, and the property that
$\rho(J^{-1}(0))\subset \gamma^{-1}(0).$
\endremark

We  now restrict to a smooth component of $J^{-1}(0)/H$ and for 
this purpose, we consider the following open submanifold of $\fg$:
$${\Cal U} =\{\,\xi\in \fg \mid {\xi}_{-\alpha_i} = (\xi, e_{\alpha_i})
\neq 0, \quad i=1,\ldots, N \,\}. \eqno(2.2.6)$$
(Note the convention in (2.2.6) is opposite to that in \c{LX2}.)
Then the $H$-action above induces a Hamiltonian
$H$-action on $TU\times {\Cal U}$ and we denote the corresponding
momentum map also by $J$ so that
$J^{-1}(0) = TU \times (\fh^{\perp}\cap {\Cal U})$.  Now recall
from \c{LX2} that there exists an 
$H$-equivariant map $g:{\Cal U}\longrightarrow H.$  Using this
map, we can identify the reduced space
$J^{-1}(0)/H = TU \times (\fh^{\perp} \cap {\Cal U}/H)$ with
$TU\times \fg_{red}$, where $\fg_{red} = \epsilon + 
\sum_{\alpha \in \Delta - \pi} {\Bbb C} e_{\alpha}$,
and $\epsilon = \sum_{j=1}^{N} e_{-\alpha_{j}}$.  Indeed, the identification
map is given by
$$ (q,p, [\xi])\mapsto (q,p, Ad_{g(\xi)^{-1}}\xi).\eqno(2.2.7)$$
Thus the natural projection $\pi_{0}:J^{-1}(0)\longrightarrow 
TU\times \fg_{red}$
is the map
$$ (q,p,\xi)\mapsto  (q,p, s= Ad_{g(\xi)^{-1}}\xi).\eqno(2.2.8)$$
Consequently, 
by Poisson reduction \c{MR}, the Poisson structure
on $TU\times \fg_{red}$ is a product structure, where the second
factor $\fg_{red}$ is equipped with the reduction (at 0) of the
Lie-Poisson structure on ${\Cal U}$.  Thus the symplectic leaves
of $TU \times \fg_{red}$ are of the form $TU \times {\Cal O}_{red}$,
where ${\Cal O}_{red} = ({\Cal O}\cap {\Cal U}\cap \fh^{\perp})/H$
and ${\Cal O}$ is an orbit in $\fg.$

\proclaim
{Proposition 2.2.6} The generic symplectic leaves in $TU \times \fg_{red}$
have dimension equal to $\dim \fg - N = 2\sum_{k=1}^{N} m_{k}.$
\endproclaim

\demo
{Proof} Clearly, the generic symplectic leaves in  $TU \times \fg_{red}$ 
correspond to generic orbits  in $\fg.$  So let ${\Cal O}$ be
a generic orbit in $\fg$ and let ${\Cal O}_{red}$ be the corresponding
reduction in $\fg_{red}$.  It is well-known that 
$\dim {\Cal O} = \dim \fg -N.$ (See, for example, \c{K2}.) Therefore, 
$\dim {\Cal O}_{red} = \dim {\Cal O} - 2N = \dim \fg - 3N.$
Consequently,
$$\eqalign{\hbox{dimension of}\,\, TU\times {\Cal O}_{red}
       & =\,\,  2N + \dim {\Cal O}_{red}\cr
       & =\,\, \dim \fg  - N.\cr}\eqno(2.2.9)$$
To complete the proof, it remains to establish the equality
$\dim \fg - N = 2\sum_{k=1}^{N} m_{k}.$  But this is just 
the assertion in (2.1.5).
\pf
\enddemo

According to the above proposition, in order to establish
the Liouville integrability of the integrable models associated
with our spin Calogero-Moser systems, we have to exhibit
$\sum_{k=1}^{N} m_{k}$ nontrivial integrals in involution which are
functionally independent on open
dense sets of the generic symplectic leaves of $TU\times \fg_{red}$.  
But as the reader will see, each of the primitive
invariants $I_{k}$ when evaluated on the Lax operators will give rise to 
$m_{k} = d_{k} -1$ such integrals. 
Hence the total number of nontrivial conserved quantities with
the required properties is exactly  $\sum_{k=1}^{N} m_{k}$.  This
explains the importance of (2.1.5).

\bigskip
\bigskip

\subhead
3. \ The rational spin Calogero-Moser systems
\endsubhead

\bigskip

The rational spin Calogero-Moser systems are associated with the
rational dynamical r-matrices with spectral parameter
$$r(q,z) = {\Omega \over z} + \sum_{\alpha\in \Delta^{\prime}}
  {1 \over \alpha(q)} e_{\alpha} \otimes e_{-\alpha}, \eqno(3.1)$$
where  $\Delta^{\prime} \subset \Delta$  is any set of roots  which
is closed with respect to addition and multiplication
by $-1,$ and $\Omega\in (S^{2}\fg)^{\fg}$ is the Casimir element
corresponding to the Killing form $(\cdot,\cdot).$  
Therefore, the Hamiltonians are given explicitly by
$${\Cal H}(q,p,\xi) = {1\over 2} \sum_{i} p_{i}^{2} - {1\over 2}
  \sum_{\alpha \in \Delta^{\prime}} \frac{\xi_{\alpha}\xi_{-\alpha}}
  {\alpha(q)^{2}}\eqno(3.2)$$
and the corresponding Lax operators are of the form
$$L(q,p,\xi)(z) = p + \sum_{\alpha \in \Delta^{\prime}} \frac{\xi_{\alpha}} 
  {\alpha(q)} e_{\alpha} + {\xi\over z}. \eqno(3.3)$$
From the homogeneity of $I_{k}$ and the form of $L(q,p,\xi)(z)$
above, $I_{k}(L(q,p,\xi)(z))$ can be expanded as
$$I_{k}(L(q,p,\xi)(z)) = \sum_{j=0}^{d_k} I_{kj}(q,p,\xi)\,z^{-j}.\eqno(3.4)$$

As the reader will see, the $I_{k1}$'s are actually identically zero
on $J^{-1}(0).$  In order to demonstrate this for all cases,
we need to establish the following lemma which is
is a refinement of (2.1.19) using just weights.

\proclaim
{Lemma 3.1}  Let $X\subset \Delta.$  Then for all $f\in I(\fg),$ $p\in \fh,$
$$\left< \partial^{n}_{p}\prod_{\a\in X} \partial^{m_{\a}} _{e_{\a}}, f
\right>  =0\eqno(3.5)$$
unless $\sum_{\a\in X} m_{\a}\, \a =0.$
\endproclaim

\demo
{Proof} For any $h\in \fh,$ it follows from (2.1.15) that
$$
\left< h\cdot \left(\partial_p^n \prod_{\a\in X} 
\partial_{e_{\a}}^{m_a}\right), f\right>=0.\eqno(3.6)
$$
Since $\fg$ acts on $S_*$ by derivation, we can use (2.1.11)
and (2.1.12) and (2.1.13) to expand the the left hand side of (3.6).
This gives 
$$\eqalign{&
\left< h\cdot \left(\partial_p^n \prod_{\a\in X} 
\partial_{e_{\a}}^{m_a}\right), f\right>\cr
= \, & n \left<\partial_p^{n-1} (h\cdot \partial_p) \prod_{\alpha\in X} 
\partial_{e_{\a}}^{m_{\a}}, f\right>+\sum_{\a\in X} m_{\a}
\left<\partial_p^n \prod_{\b\neq\a} \partial_{e_{\b}}^{m_{\b}} 
\partial_{e_{\a}}^{m_{\a}-1}(h\cdot \partial_{e_{\a}}), f\right>\cr
= \, & n \left< \partial_p^{n-1} \partial_{[h,p]} \prod_{\a\in X} 
\partial_{e_{\a}}^{m_{\a}}, f\right>+\sum_{\a\in X} m_{\a} \left<
\partial_p^n \prod_{\b\neq \a} \partial_{e_{\b}}^{m_{\b}} 
\partial_{e_{\a}}^{m_{\a}-1}\partial_{[h,e_{\a}]}, f\right>\cr
= \, & \left(\sum_{\a\in X} m_{\a}\a(h)\right)\left<\partial_p^n \prod_{\a\in X} 
\partial_{e_{\a}}^{m_{\a}}, f\right> .\cr}
$$
Therefore if $\sum_{\a\in X} m_{\a} \a\neq 0$, we must have
$\left< \partial^{n}_{p}\prod_{\a\in X} \partial^{m_{\a}} _{e_{\a}}, f
\right>  =0.$

\pf
\enddemo

\proclaim
{Proposition 3.2} For each $1\leq k\leq N$, $I_{k, d_k}(q,p,\xi)=I_{k}(\xi)$.
Moreover, for $(q,p,\xi)\in J^{-1}(0)$, we have $I_{k1}(q,p,\xi)=0.$
Hence the number of nontrivial integrals $I_{kj}(q,p, s)$ which Poisson
commute on $TU \times {\Cal O}_{red}$ is equal to
$\sum_{k=1}^{N} m_k,$ where ${\Cal O}_{red}$ is the reduction
of a generic orbit ${\Cal O}$ in ${\Cal U}.$
\endproclaim

\demo
{Proof} From the homogeneity of $I_{k}$ and the relation
$L(q,p,\xi)(z) = L(q,p,\xi)(\infty) +  {\xi\over z}$, we
have
$$I_{k}(L(q,p,\xi)(z))= I_{k}(L(q,p,\xi)(\infty)) + \ldots + 
{1\over z^{d_k}}I_{k}(\xi)$$
from which it is immediate that $I_{k, d_k}(q,p,\xi)=I_{k}(\xi).$
From the same expansion above and the definition of $I_{k1}$, we also have
$$\eqalign{
         & I_{k1}(q,p,\xi)\cr
     =\, & \lim_{z\to \infty}\, z[I_{k}(L(q,p,\xi)(z))- I_{k}(L(q,p,\xi)(\infty))]
       \cr
     =\, & {d\over dt}{\Big|_{t=0}} I_{k}(L(q,p,\xi)(\infty) + t\xi )\cr
      =\, &(\delta I_{k}(L(q,p,\xi)(\infty)), \xi).\cr}\eqno(3.7)
$$
Let us first consider the case where $\Delta^{\prime} = \Delta$.
For $(q,p,\xi)\in J^{-1}(0)$, it is clear that we have
$$\xi = \left[q, \sum_{\alpha\in \Delta} {\frac{\xi_{\alpha}}{\alpha(q)}}
e_{\alpha}\right].$$
Therefore, upon substituting into the above expression for $I_{k1}(q,p,\xi)$,
we find
$$\aligned
          &I_{k1}(q,p,\xi)\\
      =\, &(\delta I_{k}(L(q,p,\xi)(\infty)),[q, L(q,p,\xi)(\infty)])\\
      =\, & 0\\
\endaligned
$$
as $I_k$ is invariant.  In the other case where $\Delta^{\prime}\neq \Delta,$ 
let 
$\bar \Delta=\Delta\backslash \Delta'$ be the complement of $\Delta',$ then
$$
I_{k1}(q,p,\xi) = \left(\delta I_{k}(L(q,p,\xi)(\infty)),\sum_{\b\in \bar\Delta}
                   \xi_{\b} e_{\b}\right)\eqno(3.8)$$
because  $\left(\delta I_{k}(L(q,p,\xi)(\infty)),\sum_{\b\in\Delta^{\prime}}
                   \xi_{\b} e_{\b}\right)=0$
by the same reasoning as in the previous case.  Now it is clear that
the right hand side of (3.8) is linear in 
$\sum_{\b\in\bar\Delta} \xi_{\b} e_{\b}.$  In view of this, it
suffices to show that
$$\left(\delta I_{k}(L(q,p,\xi)(\infty)), e_{\b}\right)=0 \,\,\,
\hbox{for all}\,\,\b\in \bar\Delta .\eqno(3.9)$$
To this end, observe that
$$(\delta I_{k}(x), y) = {1\over (d_k -1)!}\langle \partial^{d_k-1}_{x}
\partial_y, I_k \rangle \eqno(3.10)$$
for all $x,y\in \fg.$   If we put $x = L(q,p,\xi)(\infty)$ and
$y = e_{\b}$ in the above expression and invoke the multinomial
expansion to calculate $\partial^{d_k-1}_{L(q,p,\xi)(\infty)},$
the result is
$$\eqalign{&
\left(\delta I_{k}(L(q,p,\xi)(\infty)), e_{\b}\right)\cr
=\,\,& \sum_{m + \sum_{\a\in\Delta^{\prime}} m_{\a} = d_k -1} 
 \frac{\prod_{\a\in \Delta^{\prime}} \left(\frac{\xi_\a}{\a(q)}\right)^{m_\a}}
 {m!\prod_{\a\in\Delta^{\prime}} m_{\a}!}
\left< \partial^{m}_{p} \prod_{\a\in\Delta^{\prime} }\partial^{m_\a}_{e_{\a}}
\partial_{e_{\b}}, I_k\right>.\cr}\eqno(3.11)$$
But for $\b\in \bar \Delta,$ we have 
$$\b + \sum_{\a\in\Delta^{\prime}} j_{\a} \a\neq 0\eqno(3.12)$$
for any choice of $j_{\a}\in \Bbb N, \a\in \Delta^{\prime}.$  Hence it
follows from Lemma 3.1 that each individual term of the sum in
(3.11) is equal to zero.   This completes the proof that $I_{k1}(q,p,\xi)=0$
for $(q,p,\xi)\in J^{-1}(0).$
On the other hand, it is a consquence of Proposition 6.8 in
Section 6 that all the other $I_{kj}$'s are not identically zero.
Finally, since $I_{k}(\xi)$ are Casimir functions for $1\leq k\leq N,$
the number of nontrivial integrals for each $k$ is $d_k-1 =m_k.$
\pf
\enddemo

\remark
{Remark 3.3} (a) Because the height function 
$\hbox{ht}\,:\Delta\longrightarrow \Bbb Z$
is not one-to-one, for this reason, we cannot conclude from (3.12) that
$\hbox{ht}\,(\b) + \sum_{\a\in\Delta^{\prime}} j_{\a}\hbox{ht}\,({\a} )\neq 0$
for $\b\in \bar\Delta,$  $j_{\a}\in \Bbb N, \a\in \Delta^{\prime}.$   This is why
it is necessary to use Lemma 3.1.
\newline
(b) For the rational spin CM systems considered in this section,
it was pointed out in \c{L2} that there exists a second
realization in the dual bundle of a coboundary dynamical
Lie algebroid.  More precisely, define $R: U\longrightarrow L(\fg,\fg)$
by
$$R(q)\xi = -\sum_{\alpha\in\Delta^{\prime}}\frac{\xi_{\alpha}}
  {\alpha(q)} e_{\alpha},\eqno(3.13)$$
then $R$ is a solution of the CDYBE.
Let $A^{*}\Omega \simeq TU\times \fg$ be the coboundary dynamical
Lie algebroid associated with $R$ and let $A\Omega\simeq TU\times \fg$
be the trivial Lie algebroid.  Then according to \c{L1},
$${\Cal R}: A^*\Omega \longrightarrow A\Omega, 
(q, p, \xi)\mapsto (q, \Pi_{\fh}\xi, -p + R(q)\xi)\eqno(3.14)$$
is a morphism of Lie algebroids.  Consequently, the dual map
${\Cal R}^*$ is an $H$-equivariant Poisson map, when the
domain and target are equipped with the corresponding Lie-Poisson
structures.  Explicitly,
$$\aligned
{\Cal R}^{*}(q,p,\xi) =& (q, -\Pi_{\fh}\xi, p - R(q)\xi)\\
                      =& (q, -\Pi_{\fh}\xi, L(q,p,\xi)(\infty)).\\
\endaligned
\eqno(3.15)
$$ 
We would like to point out that the (spectral parameter independent)
Lax operator 
$L^{\infty}(q,p,\xi) = L(q,p,\xi)(\infty)$ coming out from
this picture is of no use in proving Liouville integrability.  This is because
the number of integrals it gives is  far from sufficient.
The same remark also applies to the
Lax operators of the hyperbolic spin CM systems in \c{L1}
and the Lax operators of the spin CM systems associated
with the Alekseev-Meinrenken dynamical r-matrices \c{AM}
in \c{FP}.  
\endremark
\bigskip
\bigskip

\subhead
4. \ The trigonometric spin Calogero-Moser systems
\endsubhead

\bigskip

The trigonometric spin Calogero-Moser systems 
are the Hamiltonian systems in Definition 2.2.3 associated to the 
following trigonometric
dynamical r-matrices with spectral parameter:
$$r(q,z) = c(z)\sum_{i} x_{i}\otimes x_{i} -\sum_{\alpha\in \Delta}
  \phi_{\alpha}(q,z) e_{\alpha}\otimes e_{-\alpha} \eqno(4.1)$$
where 
$$c(z) = \cot z  \eqno(4.2)$$
and
$$
\phi_{\alpha}(q,z)=\cases -\frac{\sin(\alpha(q)+z)}{\sin\alpha(q)\sin z}, 
  & \alpha\in <\pi^{\prime}> \\
 -\frac{e^{-iz}}{\sin z} , & \alpha\in
  {\overline \pi^{\prime}}^+ \\
 -\frac{e^{iz}}{\sin z}, & \alpha\in
  {\overline \pi^{\prime}}^- .\endcases \eqno(4.3)
$$
In (4.3) above, $\pi^{\prime}$ is an arbitrary subset of
the simple system $\pi\subset \Delta$, $<\pi^{\prime}>$ is the root
span of $\pi^{\prime}$ and ${\overline
\pi^{\prime}}^{\pm}= \Delta^\pm \setminus <\pi^{\prime}>^\pm.$
Accordingly, the Lax operators are given by
$$\eqalign{
L(q,p,\xi)(z) =\, &p+ c(z)\sum_{i}\xi_{i}x_{i} -
  \sum_{\alpha\in \Delta} \phi_{\alpha}(q,z)\xi_{\alpha}e_{\alpha}\cr
   =\,& p + c(z)\xi + \sum_{\a\in \Delta} \psi_{\a}(q) \xi_{\a} e_{\a}\cr}
   \eqno(4.4)$$
where 
$$\psi_{\a}(q) = \cases   c(\alpha(q)),& \alpha\in <\pi^{\prime}> \\
                                      -i, & \alpha\in {\overline\pi^{\prime}}^+ \\
                                     +i, & \alpha\in {\overline\pi^{\prime}}^- .\endcases
                                       \eqno(4.5)$$
Hence we have a family of dynamical systems parametrized by subsets 
$\pi^{\prime}$ of $\pi$ with Hamiltonians of the form:
$$\eqalign{{\Cal H}(q,p,\xi) = &{1\over 2} \sum_{i} p_{i}^{2} - {1\over 2}
  \sum_{\alpha \in <\pi^{\prime}>} \left(\frac{1}{\sin^{2}\alpha(q)}
  -{1\over 3}\right) {\xi_{\alpha}\xi_{-\alpha}}
  -{5\over 6} \sum_{\alpha\in\Delta\setminus <\pi^{\prime}>}
  {\xi_{\alpha}\xi_{-\alpha}}\cr
   & -{1\over 3} \sum_{i} \xi^{2}_{i}.\cr}\eqno(4.6)$$
Now, from the homogeneity of $I_{k}$ and (4.4), we have the expansion
$$I_{k}(L(q,p,\xi)(z)) = \sum_{j=0}^{d_k} I_{kj}(q,p,\xi)(c(z))^{j}.\eqno(4.7)$$

\proclaim
{Proposition 4.1} For each $1\leq k\leq N,$ $I_{k, d_k}(q,p,\xi)=I_{k}(\xi).$
If in addition, $(q,p,\xi)\in J^{-1}(0)$, then the following relation holds:
$$\sum_{j \,\,\hbox{odd}} I_{kj}(q,p, \xi)\, i^{j} = 0.\eqno(4.8)$$
Therefore, the number of nontrivial integrals $I_{kj}(q,p,s), j\neq 1,$ which
Poisson commute on $TU\times {\Cal O}_{red}$ is equal to
$\sum_{k=1} m_k,$ where ${\Cal O}_{red}$ is the reduction
of a generic orbit ${\Cal O}$ in ${\Cal U}.$
\endproclaim

\demo
{Proof} As in the proof of Proposition 3.2, it is easy to show that
$I_{k, d_k}(q,p,\xi)=I_{k}(\xi)$ and this is a Casimir function for
each $k.$ To establish the relation
(4.8) for  $(q,p,\xi)\in J^{-1}(0)$, we divide into two cases.
First, consider $\pi^{\prime} =\pi.$  In this case,
we have
$$L(q,p,\xi)(\pm i\infty) = p + \sum_{\alpha\in \Delta} c(\alpha(q))\xi_{\alpha}
e_{\alpha}\mp i\xi .$$
Therefore, on using the relation
$(c(\alpha(q)) -i) e^{2i\alpha(q)} = c(\alpha(q)) + i$, we find that
$$Ad_{e^{2iq}} L(q,p,\xi)(i\infty) = L(q,p,\xi)(-i\infty).$$
As a consequence, we obtain 
$$I_{k}(L(q,p,\xi)(i\infty))= I_{k}(L(q,p,\xi)(-i\infty))$$
from which (4.8) follows upon using (4.7).  Now, consider the
case $\pi^{\prime}\neq \pi.$  We will establish (4.8) in this 
case through a limiting procedure.  For this purpose, we define
$$L_{q_0}(q,p,\xi)(z) = p + \sum_{\alpha\in \Delta} c(\alpha(q-q_0))\xi_{\alpha}e_{\alpha}
  +c(z)\xi, \quad q_0\in \fh.$$
Then as above, if 
$$I_{k}(L_{q_0}(q,p,\xi)(z)) = \sum_{j=0}^{d_k} I^{q_0}_{kj}(q,p,\xi)(c(z))^{j},$$
we have
$$\sum_{j \,\,\hbox{odd}} I^{q_0}_{kj}(q,p, \xi)\, i^{j} = 0.$$
Now, let $\omega_1,\ldots, \omega_N$ be the fundamental weights (with respect
to $\pi$).  We set
$q_0 = q_0(t) = - it\sum_{\alpha_j\notin \pi^{\prime}} H_{\omega_j}$ (cf. \c{EV}).
By using the relation
$$\alpha_{i}(H_{\omega_j}) = (\alpha_i,\omega_j) ={\frac{(\alpha_i,\alpha_j)}
  {2}}\delta_{ij},$$
we find that
$$\lim_{t\to \infty} c(\alpha(q-q_0(t)))= \psi_{\a} (q).$$
Therefore,
$$\lim_{t\to\infty}  L_{q_0(t)}(q,p,\xi)(z) = L(q,p,\xi)(z)$$
and so $I_{kj}(q,p,\xi) = \lim_{t\to\infty} I^{q_0(t)}(q,p,\xi).$ 
Hence we obtain (4.8) upon passing to the limit as $t\to \infty$
in the relation $\sum_{j \,\,\hbox{odd}} I^{q_0(t)}_{kj}(q,p, \xi)\, i^{j} = 0.$
By the same reason as in Proposition 3.2, all the $I_{kj}$'s are
not identically zero in this case.
Finally, since we can express $I_{k1}$ in terms of $I_{k3},\cdots$
through (4.8), the count follows.
\pf
\enddemo

\bigskip
\bigskip

\subhead
5. \ The elliptic spin Calogero-Moser systems
\endsubhead

\bigskip

Let $\wp(z)$ be the Weierstrass $\wp$-function with
periods $2 \omega_{1}$,$2\omega_{2} \in \Bbb C$, and let
$\sigma (z)$, $\zeta (z)$ be the related Weierstrass sigma-function and
zeta-function, respectively. 

The elliptic spin Calogero-Moser system is the spin Calogero-Moser
system assciated with the elliptic dynamical r-matrix with
spectral parameter
$$r(q,z) = \zeta(z)\sum_{i} x_{i}\otimes x_{i} -\sum_{\alpha\in \Delta}
  l(\alpha(q),z) e_{\alpha}\otimes e_{-\alpha} \eqno(5.1)$$
where
$$l(w,z) = -\frac{\sigma(w+z)}{\sigma(w)\sigma(z)}.\eqno(5.2)$$
Explicitly, the Hamiltonian is given by
$${\Cal H}(q,p,\xi)  = {1\over 2} \sum_{i} p_{i}^{2} - {1\over 2}
  \sum_{\alpha \in \Delta} \wp(\alpha(q)) {\xi_{\alpha}\xi_{-\alpha}}
   \eqno(5.3)$$
and its Lax operator is of the form
$$L(q,p,\xi)(z) =p+\zeta(z)\sum_{i}\xi_{i}x_{i} -
  \sum_{\alpha\in \Delta} l(\alpha(q),z)\xi_{\alpha}e_{\alpha}.
  \eqno(5.4)$$
From now onwards, we will restrict our attention  
to $(q,p,\xi)\in J^{-1}(0).$

\proclaim
{Proposition 5.1} For each $1\leq k \leq N$, $I_{k}(L(q,p,\xi)(z))$
is an elliptic function of $z$ with poles of order $d_{k}$ at the points
of the rank $2$ lattice
$$\Lambda = 2\omega_{1} \Bbb Z + 2\omega_{2}\Bbb Z.\eqno(5.5)$$
Hence $I_{k}(L(q,p,\xi)(z))$ can be expanded in the form
$$I_{k}(L(q,p,\xi)(z))= I_{k0}(q,p,\xi) + \sum_{j=2}^{d_{k}} \frac{(-1)^j}{(j-1)!}
I_{kj}(q,p,\xi)\wp^{(j-2)}(z).\eqno(5.6)$$
\endproclaim

\demo
{Proof} Let $\eta_i =\zeta(\omega_i), i=1,2.$
Then from $l(\alpha(q), z+2\omega_i) = e^{2\eta_{i}\alpha(q)} l(\alpha(q),z)$
and $e^{2\eta_{i}\alpha(q)} e_{\alpha}= Ad_{e^{2\eta_{i}q}} e_{\alpha},$
we have $L(q,p,\xi)(z+ 2\omega_i) = Ad_{e^{2\eta_{i}q}} L(q,p,\xi)(z),$
$i = 1,2.$
Therefore,  $I_{k}(L(q,p,\xi)(z))$ is a doubly-periodic function of $z.$
As $L(q,p,\xi)(z)$ is meromorphic with simple poles at the points of
the lattice $\Lambda = 2\omega_{1} \Bbb Z + 2\omega_{2}\Bbb Z$, it
follows from the homogeneity of $I_{k}$ that $I_{k}(L(q,p,\xi)(z))$
is an elliptic function of $z$ with poles of order $d_{k}$ at the points
of $\Lambda.$  The expansion of $I_{k}(L(q,p,\xi)(z))$ 
then follows from standard argument in the theory of elliptic
functions.
\pf
\enddemo

\proclaim
{Proposition 5.2} For each $1\leq k \leq N$, 
$I_{k, d_{k}}(q,p,\xi) =  I_{k}(\xi).$
Hence the number of nontrivial integrals $I_{kj}(q,p,s)$ which Poisson commute
on $TU\times {\Cal O_{red}}$ is equal to $\sum_{k=1}^{N} m_{k},$
where ${\Cal O}_{red}$ is the reduction
of a generic orbit ${\Cal O}$ in ${\Cal U}.$
\endproclaim

\demo
{Proof} In a deleted neighborhood of $z=0$, we have
$$l(\alpha(q),z) = -{1\over z} + \zeta(\a(q)) + \hbox{higher order terms}$$
from which it follows that 
$$L(q,p,\xi)(z) = p + {\xi\over z} + \sum_{\a\in \Delta}
\zeta(\a(q))\xi_{\a}e_{\a} +\hbox{higher order terms}.\eqno(5.7)$$
Therefore, on invoking the homogeneity of $I_{k}$, we obtain
the following expansion in a deleted neighborhood of $z=0$:
$$I_{k}(L(q,p,\xi)(z)) = {1\over z^{d_{k}}}\, I_{k}(\xi) + O(1).$$
But on the other hand, we have
$$\wp^{(j-2)}(z) = (-1)^{j} {\frac{(j-1)!}{z^{j}}} +  O(1)$$
for $j= 2,\ldots, d_k $.  Consequently, it follows from (5.6)
that we also have
$$I_{k}(L(q,p,\xi)(z))= {z^{-d_k}}\,
I_{k, d_{k}
}(q,p,\xi) +  O(1)$$
in a deleted neighborhood of $z=0.$  Comparing the two
expansions of $I_{k}(L(q,p,\xi)(z))$, the first assertion follows.
The second assertion is now obvious as none of the coefficients
in the expansion (5.6) is identically zero by Proposition
6.8.
\pf
\enddemo

\bigskip

\subhead
6. \ Functional independence of the integrals and Liouville integrability
\endsubhead

\bigskip

As the reader will see, we can establish the functional independence of the
integrals for all three cases in a uniform way.   For $(q,p,\xi)\in J^{-1}(0),$
we begin with the observation (see (3.3), (4.4) and (5.7)) that the Lax operator can be 
expressed in the following form
$$L(q,p,\xi) = p + h(z)\xi + k_{0}(q,\xi) + k_{1}(q,\xi,z)\eqno(6.1)
$$
in a deleted neighborhood of $0,$ where
$$ h(z) = \cases \frac{1}{z}, & \hbox{in the rational/elliptic case}\\
                           c(z), &\hbox{in the trigonometric case},\endcases
\eqno(6.2)$$
and
$$k_{0}(q,\xi) =\cases \sum_{\a\in\Delta^{\prime}}\frac{\xi_{\a}}{\a(q)} e_{\a}, 
                                            & \hbox{in the rational case}\\
                      \sum_{\a\in \Delta} \psi_{\a}(q) \xi_{\a} e_{\a}, &
                                    \hbox{in the trigonometric case}\\ 
                      \sum_{\a\in \Delta} \zeta(a(q)) \xi_{\a} e_{\a},& 
                                    \hbox{in the elliptic case},
                                    \endcases
                                    \eqno(6.3)$$    
and lastly,
$$k_{1}(q,\xi,z) = \cases  0, & \hbox{in the rational/trigonometric case}\\    
\sum_{i=1}^{\infty}k_{1i}(q,\xi) z^{i}, &\hbox{in the elliptic case.}\endcases
\eqno(6.4)$$
By using (2.1.9) and the above, it follows from the multinomial expansion that
$$\eqalign{&
I_{k}(L(q,p,\xi)(z))\cr
=\,\, &  \sum_{a +b+j =d_k} {\frac{1}{j! a! b!} }\left<{\partial_{p}^a}{\partial_{\xi}^{j}}(\partial_{k_0(q,\xi)}
             + \partial_{k_{1}(q,\xi,z)})^{b}, I_{k}\right> h(z)^{j}.\cr}\eqno(6.5)$$
We will split the second line of the above expression into a sum of two terms
$$I_{k}(L(q,p,\xi)(z)) = F_{k}(p,\xi,z) + R_{k}(q,p,\xi,z)\eqno(6.6)$$
where
$$F_{k}(p,\xi,z) = \sum_{a+j=d_k} \frac{1}{j!a!} \left<{\partial_{p}^a}{\partial_{\xi}^{j}}, I_{k}\right>
    h(z)^{j}\eqno(6.7)$$
and
$$R_{k}(q,p,\xi,z) =\sum \Sb a +b+j =d_k\\ b\geq 1\endSb
{\frac{1}{j! a! b!} }\left<{\partial_{p}^a}{\partial_{\xi}^{j}} (\partial_{k_0(q,\xi)}
             + \partial_{k_{1}(q,\xi,z)})^{b}, I_{k}\right> h(z)^{j}.\eqno(6.8)$$
Clearly, we have
$$F_{k}(p,\xi,z) = \sum_{j=0}^{d_k} F_{kj}(p,\xi)h(z)^j\eqno(6.9)$$
where
$$F_{kj}(p,\xi) = \frac{1}{j!(d_k-j)!} \left< \partial_{p}^{d_k-j}\partial_{\xi}^{j}, I_k\right>\eqno(6.10)$$
for each $j.$   Therefore these functions are the same in all three cases and the 
degree of $F_{kj}(p,\xi)$ in the variable $p$ is equal to $d_k-j.$  On the other hand,
$$R_{k}(q,p,\xi,z) = \sum_{j=0}^{d_k} R_{kj}(q,p,\xi)h(z)^j + R_{k}^{\prime}(q,p,\xi,z),
\eqno(6.11)$$
where $R_{k}^{\prime}(q,p,\xi,z)$ is identically zero in the rational/trigonometric case
and is given by a power series in $z$ which vanishes at $0$ in the elliptic case.   
From the formulas in (6.3), (6.4), it is clear that $R_{kj}$ is given by a different formula 
for each of the three cases.  However, these play no role in our analysis.  For us,
the only piece of information which is needed is the degree of $R_{kj}$ in the variable $p$ 
and according to (6.8) and (6.4), this is at most equal to  $d_k-j-1$  (and hence is less than 
that of $F_{kj}$).  We next turn to the definitions of the $I_{kj}$'s in (3.4), (4.7) and (5.6)
for the three cases.  By comparing these expressions with (6.6), (6.9)-(6.11), 
we find that
$$I_{kj}(q,p,\xi) = F_{kj}(p,\xi) + R_{kj}(q,p,\xi), \quad j=1,\cdots, d_k
\eqno(6.12)$$
in all three cases.   The relation also holds for $j=0$ for the rational/trigonometric
case but for the elliptic case, we have
$$I_{k0}(q,p,\xi) \equiv F_{k0}(p,\xi) + R_{k0}(q,p,\xi)\eqno(6.13)$$
where $\equiv$ means the two sides differ by a linear combination of
$I_{kj}(q,p,\xi)$ for $j\geq 4$ and even.  That this is so is due to contributions from the
constant terms in the Laurent series expansions of $\wp^{(j-2)}(z)$ on the right
hand side of (5.6) for $j\geq 4$ and even.
 
\proclaim
{Proposition 6.1}   The functional independence of $F_{kj}(p,\xi),$ $j=0,\widehat{1},\cdots, d_k,$
$k=1,\cdots, N$ on an open dense set of $\fh\times ({\Cal U}\cap \fh^{\perp})$ 
implies the 
functional independence of $I_{kj}(q,p,\xi),$
$j=0, \widehat{1},\cdots, d_k,$ $k=1,\cdots, N$ on an open dense set of
$TU\times ({\Cal U}\cap \fh^{\perp}).$
\endproclaim

\demo
{Proof}  Suppose the $I_{kj}$'s are functionally dependent.  Then
there exists an analytic function $f(u_1,\cdots,u_d)$ depending on
$d= {1\over 2}(\hbox{dim}\,\fg + N)$ variables such that $f(I_{kj}(q,p,\xi))=0.$
Fix a point $q=q_0\in U$, then  $f(I_{kj}(q_0,p,\xi))=0$ 
is a functional dependence relation among the polynomials 
$I^{q_0}_{kj}(p,\xi):=I_{kj}(q_0,p,\xi)$
in $p$ and $\xi.$   Since analytic dependence implies algebraic dependence 
for polynomials
(see, for example, \c{W} and the references therein),  we can assume that
$f$ is a polynomial in the variables $u_1, \cdots, u_d.$  Now the
highest order term in $p$ in the expression $f(I_{kj}(q_0,p,\xi))$ is of 
the form $g(F_{kj}(p,\xi))$ for a summand $g$ of $f,$ since for each 
monomial $I_{10}^{n_1}\cdots I_{Nd_N}^{n_d},$ the highest order term in
$p$ is given by $F_{10}^{n_1}\cdots F_{Nd_N}^{n_d}.$  Furthermore, since
$f$ is not identically zero, neither is $g.$
But $f(I_{kj}(q,p,\xi))=0$ implies $g(F_{kj}(p,\xi))=0,$ hence
the $F_{kj}$'s are functionally dependent.
\pf
\enddemo

In what follows, we will establish the functional independence 
of $F_{kj}(p,\xi),$ $j=0,\widehat{1},\cdots, d_k,$ $k=1,\cdots, N$ on an
open dense set of $\fh\times ({\Cal U}\cap \fh^{\perp}).$
The following is a lemma which is very useful in some of
our calculations.

\proclaim
{Lemma 6.2}  Let  $f\in I(\fg),$  then for all $x,y,z\in \fg$,
and all $m,n \geq 0,$ we have
$$
\langle \partial_x^m\partial_{[x,y]}\partial_z^n,f\rangle
=\frac{n}{m+1}\langle \partial_x^{m+1}\partial_{[y,z]}\partial_z^{n-1},f\rangle
,\eqno(6.14)$$
where by convention the right hand side of the formula is zero when
$n=0.$ 
\endproclaim

\demo
{Proof} By using (2.1.13), (2.1.12) back and forth and (2.1.11), we
find for $n\geq 1$ that
$$\align
 \langle \partial_x^{m} \partial_{[x,y]}\partial_z^n, f\rangle = &
-\langle \partial_x^{m} (y\cdot \partial_{x})\partial^n, f\rangle\\
= & -\frac{1}{m+1} \langle (y\cdot \partial_x^{m+1})\partial_z^n,f\rangle\\
= &-\frac{1}{m+1} \langle y\cdot (\partial_x^{m+1}\partial_z^n),f\rangle
   +\frac{1}{m+1} \langle \partial_x^{m+1}(y\cdot \partial_z^n),f\rangle\\
= & \frac{n}{m+1}\langle \partial_x^{m+1}(y\cdot \partial_z)\partial_z^{n-1},f
    \rangle\\
= & \frac{n}{m+1} \langle \partial_x^{m+1}\partial_{[y,z]}\partial_z^{n-1},f
    \rangle
\endalign
$$
where we have used (2.1.15) in additon to the ``power rule''
in going from the third line to the
fourth line.  When $n=0,$ the calculation stops in the second
line for we can invoke (2.1.15) to conclude that the resulting
expression is equal to zero.
\pf
\enddemo

\proclaim
{Proposition 6.3} For all $1\leq k\leq N,$ $(p,\xi)\in \fh\times ({\Cal U}\cap
\fh^{\perp}),$
\newline
(a) $F_{k0}(p,\xi) = I_{k}(p),$
\newline
(b) $F_{k1}(p,\xi) = 0,$
\newline
(c) $F_{k,d_k}(p,\xi) = I_{k}(\xi).$
\endproclaim

\demo
{Proof} The assertions in (a) and (c) are obvious.  For (b), we use the 
representation in 
(6.l0) together with the fact that $\partial_{\xi}$  has no weight
zero part for $\xi\in \fh^{\perp}.$  The assertion therefore
is a consequence of (2.1.19).
\pf
\enddemo

In order to set up our calculation, we will arrange the
variables and the functions $F_{kj}$ in some definite
order.  Note 
that for each $1\leq k\leq N,$
the number of $F_{kj}(p,\xi)$'s with $j\neq 1$
is equal to $d_k.$  Therefore we have a partition given by
the sequence
$$h = d_N\geq d_{N-1}\geq \cdots \geq d_1=2.\eqno(6.15)$$
Since $d_k = m_k +1,$ it is easy to show from Theorem 2.1.2
that the above sequence is conjugate to the partition
$$N=b_0=b_1\geq b_2\geq \cdots \geq b_{h-1} =1.\eqno(6.16)$$
The ordering of the $F_{kj}$'s which we will use is the following:
$$F_{10},\cdots, F_{N0};F_{12},\cdots, F_{N2};
 F_{n-b_2 +1,3},\cdots,F_{N3};\cdots;F_{N, d_N}.\eqno(6.17)$$
Clearly, for each value of $j\geq 2,$ the number of functions
in each group $\{F_{kj}\}$ is precisely $b_{j-1}$ from our
discussion above.
Now for each $1\leq j\leq h-1,$ let us denote the roots with
height equal to $j$ by $\a_{j,i}, i=1,\cdots, b_j.$
We  will order the variables as depicted in the following:
$$p_1,\cdots, p_N; \xi_{\a_1},\cdots, \xi_{\a_N};
\xi_{\a_{2,1}},\cdots, \xi_{\a_{2,b_2}};\cdots ;\xi_{\a_{h-1,1}}.\eqno(6.18)
$$

\proclaim
{Theorem 6.4} The functions $F_{kj}(p,\xi),$ $j=0,\widehat{1},\cdots, d_k,$
$k=1,\cdots, N$ are functionally independent on an open dense set of 
$\fh\times ({\Cal U}\cap \fh^{\perp}).$ 
\endproclaim

To prove this assertion, we will compute the coefficient of
$$dp_1\wedge\cdots\wedge dp_N\wedge d\xi_{\a_1}\wedge\cdots \wedge d\xi_{\a_N}
\wedge \cdots \wedge d\xi_{\a_{h-1,1}}$$ 
in the expression for
$$dF_{10}\wedge \cdots \wedge dF_{N0}\wedge dF_{1,2}\wedge\cdots\wedge dF_{N2}
\wedge \cdots \wedge dF_{Nd_N}$$
at the points of  $\fh\times(\epsilon + \frak n),$ 
where $\epsilon$
is as in Section 2.2 and $\frak n$ is the nilpotent subalgebra
$\sum_{\a\in \Delta^+} \fg_{\a}.$  Note that the choice of $\epsilon + \frak n$
follows Kostant in \c{K2}.  Indeed, if $e_+ =\sum_{\a\in \pi} c_{\a}e_{\a},
c_{\a}\neq 0$ for all $\a\in \pi,$ then Kostant showed that
the $N$-dimensional plane 
$\frak o =\epsilon + \fg^{e_+}\subset \epsilon + \frak n$ is
a global cross-section of the generic orbits in $\fg$ in the sense
that each such orbit intersects $\frak o$ at precisely one point
and no two distinct points in $\frak o$ are conjugate.  This
is the reason why it suffices to consider $\fh\times (\epsilon + \frak n).$

\remark
{Remark 6.5} For $\fg = sl(N+1,\Bbb C),$ the generic
orbits can be characterized as those orbits through matrices 
whose characteristic polynomial and minimal polynomial coincide.
In this case, we can take $\frak o$ to be the set of companion matrices
and the result of Kostant which we quoted above is well-known in
matrix theory. (See, for example, \c{HJ}.)
\endremark

The computation which we referred to above will be achieved in a 
sequence of propositions.  First of all,
the coefficient which we want to compute is the determinant of
a square (block) matrix $D$ of partial derivatives whose diagonal
blocks are given by
$$D_0 = \left(\frac{\partial F_{l0}}{\partial p_i}\right)_{l,i=1}^{N},\,\,
D_j = \left(\frac{\partial F_{N-b_j +l, j+1}}{\partial \xi_{\a_{j,i}}}
\right)_{l,i=1}^{b_j}, \quad j=1,\cdots, h-1,\eqno(6.19)$$
in that order.

\proclaim
{Proposition 6.6} At the points $(p,\xi)\in \fh\times(\epsilon + \frak n),$
\newline
(a) $F_{k0}$ does not depend on $\xi_{\a}$ for all $\a\in \Delta^+,$
\newline
(b) for $j\geq 2,$ $F_{kj}$ does not depend on $\xi_{\a}$ for those
$\a$ with $\hbox{ht}\,(\a)\geq j,$ and it depends linearly on
$\xi_{\a}$ for those $\a$ with $\hbox{ht}\,(\a) =j-1.$
\newline
(c)  $D$ is block lower-triangular, i.e.,
$$D = \pmatrix 
D_0 & & & & \\
& D_1 & & 0 & \\
 & & D_2 & & \\
 & * & & \cdots & \\
 & & & & D_{h-1}
\endpmatrix ,\eqno(6.20)
$$
and the square blocks $D_j$ defined in (6.19) depend only
on $p.$
\endproclaim

\demo
{Proof}  (a)  This is just a consequence of  Proposition 6.3 (a).
\newline
(b) This part follows from weight consideration.  Apply (6.10)
with $\xi = \epsilon + \xi^+ = \epsilon + \sum_{\a\in \Delta^+}
\xi_{\a} e_{\a}$ and apply the binomial expansion to 
calculate $(\partial_{\epsilon} + \partial_{\xi^+})^j,$ we have
$$F_{kj}(p,\xi) \equiv \langle \partial_{p}^{d_k-j }\partial_{\epsilon}^{j}, I_k\rangle
    + j \sum_{\a\in\Delta^+} \xi_{\a} \langle \partial_{p}^{d_k-j}\partial_{\epsilon}^{j-1}
    \partial_{e_{\a}}, I_k\rangle +  O(\xi^2).\eqno(6.21)$$
 Here the notation $a\equiv b$ is a shorthand for $a= \lambda b$ for some
 $\lambda\neq 0$ and we will henceforth use this shorthand.
  On the other hand,  the reminder term $O(\xi^2)$ involves
 terms which are at least quadratic in the components of $\xi^+.$
 From (2.1.17) and (2.1.18), $\partial_{p}^{d_k-j}$ has weight $0$
 while $\partial_{\epsilon}^{j}$ has weight $-j$.  Therefore the 
 first term in (6.21) is zero by (2.1.19).  If  $\hbox{ht}\,(\a)\geq j,$
 then $\partial_{\epsilon}^{j-1}\partial_{e_{\a}}$ has weight
 strictly bigger than $0$ and therefore the corresponding term
$\langle \partial_{p}^{d_k-j}\partial_{\epsilon}^{j-1}
    \partial_{e_{\a}}, I_k\rangle$ in (6.21)
 is zero by (2.1.19).  On the other hand, if $\hbox{ht}\,(\a)=j-1,$ the operator
 $\partial_{\epsilon}^{j-1}\partial_{e_{\a}}$ 
 has weight $0$ and therefore the corresponding $\xi_{\a}$ appears linearly
 in $F_{kj}.$  Finally, it is clear that the term $O(\xi^2)$
does not depend on $\xi_{\a}$ for $\a$ with height greater
or equal to $j.$  This completes the argument.
\newline
(c) This immediately follows from the assertions in (a), (b) and
(6.19).
\pf
\enddemo

We next compute the values of the determinants  $|D_j|, j=0,\cdots, h-1.$
For this purpose, we have to study the diagonal blocks
of $D$ more closely.

\proclaim
{Proposition 6.7}  At the points $(p,\xi)\in \fh\times(\epsilon + \frak n),$
the following properties hold.
\newline
(a) For $1\leq l,i\leq b_j,$ the element $D_j(l,i)$ of $D_j$ in the
$(l,i)$ position has degree $d_{N-b_j +l} -j-1$ in $p.$
\newline
(b) The first $b_j-b_{j+1}$ rows of $D_j$ are constants.
(When $b_{j+1} =b_j,$ this just means that there are no constant rows.)
Indeed, when $b_j-b_{j+1} >0,$ we have the formula
$$D_j(l,i)\equiv \langle \partial_{\epsilon}^j \partial_{e_{\a_{j,i}}},I_{N-b_j+l}\rangle,\hbox{ for }\,
1\leq l\leq b_{j}-b_{j+1}.\eqno(6.22)$$

\endproclaim

\demo
{Proof}
(a) For $j=0,$ the assertion is clear because $F_{l0}$ is homogeneous
of degree $d_{l}$ in $p$ by Proposition 6.3 (a).  For $j\geq 2,$
it follows from (6.10), (6.19) and (6.21) that
$$\eqalign{D_j(l,i) = & \,\frac{\partial F_{N-b_j +l, j+1}} 
                        {\partial \xi_{\a_{j,i}}}\cr
               \equiv & \,\langle \partial_p^{d_{N-b_j+l}-j-1} 
                        \partial_{\epsilon}^{j} 
                        \partial_{e_{\a_{j,i}}},I_{N-b_j+l}\rangle .\cr}
\eqno(6.23)
$$
Hence the degree of $D_j(l,i)$ in $p$ is $d_{N-b_j+l}-j-1.$
\newline
(b) If $b_j -b_{j+1} >0,$ we have $m_k =j$ for
$N-b_j +1 \leq k\leq N-b_{j+1}$ from Theorem 2.1.2 (a) which implies
$d_{N-b_j +l} = j+1$ for $l=1,\cdots, b_j-b_{j+1}.$
Thus $D_j(l,i)$ is of degree $0$ in $p$ for $l=1,\cdots, b_j-b_{j+1}$
from (6.23), i.e., they are constants.
\pf
\enddemo 

\proclaim
{Proposition 6.8}  Let  $\Delta^{+}_{j}$ denote the set of 
positive roots of height $j.$  Then on $\fh,$ we have
the recursion relations:
\newline
(a) $|D_1|\,\prod_{i=1}^{N} \a_i \equiv |D_0|$, 
\newline
(b) $|D_{j}|\,\prod_{\a\in \Delta^{+}_{j}} \a \equiv |D_{j-1}| \,\,\,\, \hbox{for}
\,\,j\geq 2,$
\newline
where the proportionality constants in (a) and (b) are independent of 
$p\in\fh.$
\newline
Therefore 
$$|D_j|(p)\equiv \prod_{\hbox{ht}\,(\a)>j}\a(p),\,\,j=0,1,\cdots,h-1
\eqno(6.24)$$
with the convention that $|D_{h-1}|(p)\equiv 1.$
Hence $|D_j|(p)\neq 0$ for $p\in \fh^{\prime},$
$j=0,1,\cdots,h-1,$  where $\fh^{\prime}$ is the open, dense
set of regular points of $\fh.$
\endproclaim

\demo
{Proof} It is a classical result that 
$|D_0|(p) \equiv \prod_{\a\in \Delta^{+}} \a(p)$ 
and the regular points of $\fh$ are precisely those points
where $|D_0|(p)\neq 0.$ (See \c{S} and \c{K2}.)
Therefore, if we can establish the recursion relations,
it will follow from this result that
$|D_j|(p)\neq 0$ for $p\in \fh^{\prime},$
$j=0,1,\cdots,h-1.$
\newline
(a) Let $H_i = H_{\a_i}, i=1,\cdots, N.$  Since the $H_i$'s form
a basis of $\fh,$ the determinant $|D_0|$ in (6.19) can be
computed in this basis up to a nonzero scale. That is,
$$|D_0|(p) = \Bigl|\bigl(\langle \partial_p^{d_l-1} \partial_{x_i}, I_l\rangle\bigr)_{l,i}\Bigr|\equiv \Bigl|\bigl(\langle \partial_p^{d_l-1} \partial_{H_i}, I_l\rangle\bigr)_{l,i}\Bigr|.$$
But from (6.23), (6.14) and
the relation $[e_{\a_i}, \epsilon] =H_i,$ we have
$$\eqalign{
\a_i(p) D_1(l,i)(p)\equiv \, & \a_i(p)\langle \partial_p^{d_l-2} 
  \partial_{\epsilon}\partial_{e_{\a_i}}, I_l\rangle\cr
=\,& \langle \partial_p^{d_l-2} \partial_{\epsilon}\partial_{[p,e_{\a_i}]}, I_l
   \rangle\cr
\equiv \,& \langle \partial_p^{d_l-1} \partial_{[e_{\a_i},\epsilon]},I_l
   \rangle\cr
=\,&\langle \partial_p^{d_l-1} \partial_{H_i},I_l\rangle .\cr}
$$
Hence the formula follows from the property of determinants.
\newline
(b) Consider the root vector $e_{\a_{j,i}}.$  Clearly
we have $e_{\a_{j,i}}\in \fg^{(j)}$ and $\epsilon\in \fg^{(-1)}.$ (See
the definition at the end of Section 2.1.)  Therefore
$[\,e_{\a_{j,i}}, \epsilon\,]\in \fg^{(j-1)}$ by (2.1.20).
Hence we can write
$$[\,e_{\a_{j,i}}, \epsilon\,] = \sum_{n=1}^{b_{j-1}} a_{j,n,i} e_{\a_{j-1,n}},
\eqno(6.25)$$
where the coefficients on the right hand side are not all
zero.  Indeed, it follows from 
(4.4.3) and the proof of Proposition 19 in \c{K1} that
$\hbox{ker}\,(ad\,\epsilon)\cap \frak n =0$
and therefore the $b_{j-1}\times b_j$ matrix $A_j =(a_{j,n,i})_{n,i}$
is of full rank.
Now, by making use of the formula for $D_j(l,i)$ in (6.23),
it follows by applying (6.14) and (6.25) that
$$\eqalign{
\a_{j,i}(p) D_j(l,i)(p)\equiv\, & \a_{j,i}(p)\,\langle \partial_p^{d_{N-b_j+l}-j-1} 
\partial_{\epsilon}^{j} \partial_{e_{\a_{j,i}}},I_{N-b_j+l}\rangle\cr
=\,& \langle \partial_p^{d_{N-b_j+l}-j-1} \partial_{\epsilon}^j
 \partial_{[p,e_{\a_{i,i}}]}, I_{N-b_j+l}\rangle\cr
\equiv \, & \langle \partial_p^{d_{N-b_j+l}-j} \partial_{\epsilon}^{j-1} 
\partial_{[e_{\a_{j,i}}, \epsilon]},I_{N-b_j+l}\rangle \cr
=\,&\sum_{N=1}^{b_{j-1}} a_{j,n,i} \langle \partial_p^{d_{N-b_j+l}-j} \partial_{\epsilon}^{j-1} \partial_{e_{\a_{j-1,n}}},I_{N-b_j+l}\rangle\cr
=\, &\sum_{n=1}^{b_{j-1}} a_{j,n,i} D_{j-1}(l+b_{j-1}-b_j,n)(p).\cr}\eqno(6.26)
$$
We now divide the proof into two cases.
\newline
Case 1. $b_{j-1} =b_j$
\smallskip
In this case, we have
$$D_j(p)\, \hbox{diag}\,(\a_{j,1}(p),\cdots, \a_{j,b_j}(p)) 
\equiv D_{j-1}(p) A_j\eqno(6.27)$$
from (6.26) above and the matrix $A_j$ is invertible.  Therefore, when we take 
the determiant of both sides of (6.27), we obtain the desired formula.
\newline
Case 2. $b_{j-1} > b_j$
\smallskip
In this case, (6.26) can be rewritten as
$$D_j(p)\, \hbox{diag}\,(\a_{j,1}(p),\cdots, \a_{j,b_j}(p))
\equiv D^{\prime}_{j-1}(p)A_j\eqno(6.28)$$
where $D^{\prime}_{j-1}(p)$ is the  $b_j\times b_{j-1}$ submatrix
of $D_{j-1}(p)$ obtained by deleting its first $b_{j-1}-b_j$
rows.  Now, recall that the first $b_{j-1} -b_j$ rows of $D_j$ are constants in 
this case by Proposition 6.7 (b).
Consequently, for $1\leq l\leq b_{j-1}-b_j,$ it follows by using (6.22) and by 
reversing the
steps in the kind of calculation in (6.26) that
$$\eqalign{
\sum_{n=1}^{b_{j-1}} a_{j,n,i} D_{j-1}(l,n)(p) \equiv \, & \sum_{n=1}^{b_{j-1}} a_{j,n,i}
\langle \partial_{\epsilon}^{j-1}\partial_{e_{\a_{j-1,n}}},I_{N-b_{j-1} +l }\rangle\cr
=\, & \langle\partial_{\epsilon}^{j-1}\partial_{[e_{\a_{j,i}}, \epsilon]},
I_{N-b_{j-1}+l} \rangle\cr
=\, & 0\cr}\eqno(6.29)
$$
where we have used (6.14) in the $n=0$ case and (6.25) in going from the
first line to the second line.  By combining (6.28) and (6.29),
we conclude that
$$\pmatrix   0\\ D_j(p)\, \hbox{diag}\,(\a_{j,1}(p),\cdots, \a_{j,b_j}(p))
\endpmatrix
= D_{j-1}(p) A_j.\eqno(6.30)$$
But since the $b_{j-1}\times b_j$ matrix $A_j$ has full rank,
we can extend it to an invertible $b_{j-1}\times b_{j-1}$ matrix
$\widetilde {A_j}$ by adjoining 
$b_{j-1}-b_j$ column vectors from the canonical basis of $\Bbb C^{b_{j-1}}$  
on the right hand side of $A_j.$ 
In this way, we obtain from (6.30) that
$$\pmatrix  0 & \# \\
   D_j(p)\, \hbox{diag}\,(\a_{j,1}(p),\cdots, \a_{j,b_j}(p)) & *\endpmatrix
   = D_{j-1}(p) \widetilde {A_j}.\eqno(6.31)$$
Therefore, on taking the determinants of both sides of (6.31), we again 
obtain the desired formula.
\pf
\enddemo

This proves Theorem 6.4 as
$$\eqalign{
|D|(p) \equiv \, & \prod_{j=0}^{h-2} \left(\prod_{ht\,(\a)>j}\a(p)
\right)
\cr
=\, & \prod_{\a\in \Delta^+} \a(p)^{ht\,(\a)} \neq 0\cr}
\eqno(6.32)$$
for $p\in \fh^{\prime}.$

As a consequence of Theorem 6.4 and Propositon 6.1, we obtain the
following corollary.

\proclaim
{Corollary 6.9}  The Poisson commuting integrals $I_{kj}(q,p,\xi),$
$j=0, \widehat{1},\cdots, d_k,$ $k=1,\cdots, N$ on 
$TU\times ({\Cal U}\cap \fh^{\perp})$ are functionally
independent on an open dense set of
$TU\times ({\Cal U}\cap \fh^{\perp}).$
\endproclaim

Finally we are ready to state the main theorem of this work.

\proclaim
{Theorem 6.10} The reduction of the rational, trigonometric and
elliptic spin Calogero-Moser systems to 
$J^{-1}(0)/H \simeq TU\times \fg_{red}$ are Liouville integrable on the
generic symplectic leaves of $TU\times \fg_{red}.$
\endproclaim

\demo
{Proof} With the identification $J^{-1}(0)/H \simeq TU\times \fg_{red},$
the conserved quantitites in involution are given by 
$I_{kj}(q,p,s),$ where $s\in \fg_{red}.$  Therefore
the number of nontrivial integrals required for Liouville
integrability is exactly one-half the dimension of the
generic symplectic leaves of $TU\times \fg_{red}$
for each of the three cases. (See Proposition 3.2, 4.1
and 5.2.) Finally, the functional independence of 
the integrals follows from Corollary 6.9 above.
\pf
\enddemo


\bigskip
\bigskip


\bigskip
\newpage

\Refs
\widestnumber\key{RSTS}

\ref\key{AM} 
\by Alekseev, A. and Meinrenken, E.
\paper Clifford algebras and the classical dynamical
Yang-Baxter equation
\jour Math. Res. Lett.\vol 10\yr 2003\pages 253-268
\endref

\ref\key{B}
\by Bott, R.
\paper An application of Morse theory to the
topology of Lie groups
\jour Bull. Soc. Math. Fr.\vol 84\yr 1956\pages 251-258
\endref

\ref\key{BAB1}
\by Billey, E., Avan, J. and Babelon, O.
\paper The r-matrix structure of the Euler-Calogero-Moser model
\jour Phys. Lett. A\vol 186\yr 1994\pages 263-271
\endref

\ref\key{BAB2}
\by Billey, E., Avan, J. and Babelon, O.
\paper  Exact Yangian symmetry in the classical Euler-Calogero-Moser
model
\jour Phys. Lett. A\vol 188\yr 1994\pages 263-271
\endref

\ref\key{C}
\by Chevalley, C.
\paper Invariants of finite groups generated by reflections
\jour Amer. J. Math.\vol 77\yr 1955\pages 778-782
\endref

\ref\key{CM}
\by Collingwood, D. and McGovern, W.
\book Nilpotent orbits in semisimple Lie algebras
\publ Van Nostrand Rheinhold\publaddr New York
\yr 1993
\endref

\ref\key{DLT}
\by Deift, D., Li, L.-C. and Tomei, C.
\paper Matrix factorization and integrable systems
\jour Comm. Pure Appl. Math.\vol 42\yr 1989\pages 443-521
\endref

\ref\key{EV}
\by Etingof, P. and Varchenko, A.
\paper Geometry and classification of solutions of the classical dynamical
Yang-Baxter equation
\jour Commun. Math. Phys.\vol 192\yr 1998 \pages77-120
\endref 

\ref\key{GH}
\by Gibbons, J. and Hermsen, T.
\paper A generalization of the Calogero-Moser systems
\jour Physica D\vol  11D \yr  1984 \pages 337-348
\endref

\ref\key{F}
\by Felder, G.
\paper Conformal field theory and integrable systems associated to elliptic
curves
\jour  Proc. ICM Zurich, Birkh\"auser, Basel\yr 1994\pages1247--1255
\endref

\ref\key{FP}
\by Feher, L. and Pusztai, G.
\paper Spin Calogero-Moser models obtained from dynamical
r-matrices and geodesic motion
\jour Nucl. Phys. B\vol 734\yr  2006\pages 304-325
\endref

\ref\key{HH}
\by Ha, Z.N.C. and Haldane, F.D.M.
\paper On models with inverse-square exchange
\jour Phys. Rev. B\vol 46\yr 1992\pages 9359-9368
\endref

\ref\key{HJ}
\by Horn, R. and Johnson, C.
\book Matrix analysis
\publ Cambridge University Press\publaddr Cambridge\yr 1985
\endref


\ref\key{K1}
\by Kostant, B.
\paper The principal three-dimensional subgroup and the
Betti numbers of a complex simple Lie group
\jour Amer. J. Math.\vol 81\yr 1959\pages 973-1032
\endref

\ref\key{K2}
\by Kostant, B.
\paper Lie group representations on polynomial rings
\jour Amer. J. Math.\vol 85\yr 1963\pages 327-404
\endref

\ref\key{L1}
\by Li, L.-C.
\paper A family of hyperbolic spin Calogero-Moser systems and
the spin Toda lattices 
\jour Comm. Pure Appl. Math.\vol 57\yr 2004\pages 791-832
\endref

\ref\key{L2}
\by Li, L.-C.
\paper  A class of integrable spin Calogero-Moser systems II:exact solvability
\jour IMRP Int. Math. Res. Pap. 2006, Art. ID 62058\pages 53 pp
\endref 

\ref\key{L3}
\by Li, L.-C.
\paper Poisson involutions, spin Calogero-Moser systems associated with
symmetric Lie subalgebras and the symmetric space spin Ruijsenaars-Schneider
models
\jour Commun. Math. Phys.\vol 265\yr 2006\pages 333-372
\endref

\ref\key {LX1}
\by  Li, L.-C. and Xu, P.
\paper Spin Calogero-Moser
systems associated with simple Lie algebras
\jour C. R. Acad. Sci. Paris, S\'erie I\vol 331\yr 2000\pages 55--60
\endref

\ref\key{LX2}
\by Li, L.-C. and Xu, P.
\paper A class of integrable spin Calogero-Moser systems
\jour  Commun. Math. Phys. \vol 231 \yr 2002 \pages 257-286
\endref

\ref\key{MR}
\by  Marsden, J. and Ratiu, T.
\paper Reduction of Poisson manifolds
\jour Lett. Math. Phys. \vol 11\yr 1986 \pages161--169 
\endref

\ref\key{MP}
\by Minahan, J.A. and Polychronakos, A.
\paper Interacting Fermion systems from two-dimensional QCD
\jour Phys. Lett. B\vol 326\yr 1994\pages 288-294
\endref

\ref\key{OP}
\by Olshanetsky, M. and Perelomov, A.M.
\paper Completely integrble Hamiltonian systems connected with
semisimple Lie algebras
\jour Invent. Math.\vol 37\yr 1976\pages 93-108
\endref

\ref\key{OR}
\by Ortega, J.-P. and Ratiu, T.
\paper Singular reduction of Poisson manifolds
\jour Lett. Math. Phys. \vol 46\yr 1998 \pages 359-372
\endref

\ref\key{P}
\by Polychronakos, A.
\paper Calogero-Moser systems with noncommutative spin interactions
\jour Phys. Rev. Lett.\vol 89\yr 2002\pages 126403
\endref

\ref\key{Pech}
\by Pechukas, P.
\paper Distribution of energy eigenvalues in the irregular spectrum
\jour Phys. Rev. Lett.\vol 51\yr 1983\pages 943-946
\endref

\ref\key{RSTS}
\by Reyman, A. and Semenov-Tian-Shansky, M.
\paper Group-theoretical methods in the theory of finite-dimensional
integrable systems 
\inbook Dynamical Systems VII, Encyclopaedia of Mathematical Sciences,
\vol 16
\eds V.I. Arnold and S.P. Novikov
\publ Springer-Verlag \yr 1994\pages 116-225
\endref

\ref\key{ST}
\by Shephard, G.C. and Todd, J.A.
\paper Finite unitary reflexion groups
\jour Can. J. Math.\vol 6\yr 1954\pages 274-304
\endref

\ref\key{S}
\by Steinberg, R.
\paper Invariants of finite reflection groups
\jour Can. J. Math.\vol 12\yr 1960\pages 616-618
\endref

\ref\key{V}
\by Varadarajan, V.S.
\paper On the ring of invariant polynomials on a simple Lie algebra
\jour Amer. J. Math.\vol 90\yr 1968\pages 308-317
\endref

\ref\key{W}
\by Whitney, H.
\book Complex analytic varieties
\publ Addison-Wesley \publaddr Reading, Mass.-London-Don Mills, Ont.
\yr 1972
\endref

\ref\key{Wo}
\by Wojciechowski, S.
\paper An integrable marriage of the Euler equations with the Calogero-Moser
systems
\jour Phys. Lett. A\vol 111\yr 1985\pages 101-103
\endref

\ref\key{Y}
\by Yukawa, T.
\paper New approach to the statistical properties of energy levels
\jour Phys. Rev. Lett.\vol 54\yr  1985\pages 1883-1886
\endref

\endRefs
\enddocument